\documentclass[conference]{IEEEtran}
\IEEEoverridecommandlockouts

\usepackage{cite}
\usepackage{amsmath,amssymb,amsfonts}
\usepackage{algorithmic}
\usepackage{graphicx}
\usepackage{textcomp}
\usepackage{xcolor}
\usepackage{float}
\usepackage{hyperref}
\usepackage{booktabs}
\usepackage{balance}

\def\ninept{\def\baselinestretch{1.05}\let\normalsize\small\normalsize}

\def\BibTeX{{\rm B\kern-.05em{\sc i\kern-.025em b}\kern-.08em
    T\kern-.1667em\lower.7ex\hbox{E}\kern-.125emX}}

\begin{document}
\ninept

\title{Semi-intrusive audio evaluation: \\Casting non-intrusive assessment as a multi-modal text prediction task\\
\thanks{GPT-4o was used for editing and grammar enhancement.}
}

\author{\IEEEauthorblockN{Jozef Coldenhoff}
\IEEEauthorblockA{
\textit{Logitech Europe S.A.}\\
Lausanne, Switzerland \\}
\and
\IEEEauthorblockN{Milos Cernak}
\IEEEauthorblockA{
\textit{Logitech Europe S.A.}\\
Lausanne, Switzerland \\}
}

\maketitle

\begin{abstract}

Human perception has the unique ability to focus on specific events in a mixture of signals -- a challenging task for existing non-intrusive assessment methods. In this work, we introduce semi-intrusive assessment that emulates human attention by framing audio assessment as a text-prediction task with audio-text inputs. To this end, we extend the multi-modal PENGI model through instruction fine-tuning for MOS and SNR estimation. For MOS, our approach achieves absolute Pearson correlation gains of 0.06 and 0.20 over the re-trained MOSRA model and the pre-trained PAM model, respectively. We further propose a novel SNR estimator that can focus on a specific audio source in a mixture, outperforming a random baseline and the fixed-prompt counterpart. Our findings suggest that semi-intrusive assessment can effectively capture human-like selective listening capabilities. Samples are available at \url{https://jozefcoldenhoff.github.io/semi-intrusive-assessment}.
\end{abstract}


\section{Introduction}
Audio assessment entails quantifying the degree of impairment in an audio signal. In terms of perceptual audio \textit{quality}, the gold standard for assessment remains conducting human listening tests through, for example, the ITU-T recommendation p.800 \cite{p800rec} for speech, or its more general audio counterpart in ITU-R Recommendation BS.1284 \cite{BS_1284}. However, given the costly nature of conducting human listening tests, objective measures have been developed.

Early works largely concentrated on intrusive methods, where degraded signals are evaluated by comparing them to their clean references. Traditional metrics like signal-to-noise ratio and log spectral distance \cite{Dimolitsas_1989} were commonly employed for this task. More recent developments, such as VISQOL \cite{chinen2020visqolv3opensource}, PESQ \cite{pesq}, and POLQA~\cite{beerends2013perceptual}, have introduced more sophisticated systems designed explicitly for intrusive quality evaluation.

Still, the reliance on a matching clean reference signal limits the scope of their application. Therefore, non-intrusive methods have been developed, with many relying on deep learning to perform audio assessment. Methods such as TorchAudio-Squim \cite{kumar2023torchaudiosquimreferencelessspeechquality} for non-intrusive SI-SDR estimation or other early works for mean opinion score estimation (MOS) such as AutoMOS \cite{patton2016automoslearningnonintrusiveassessor} for speech synthesis, and Quality-Net \cite{fu2018qualitynetendtoendnonintrusivespeech} for speech processed by a denoising algorithm aim to overcome the reliance on a clean reference.

However, methods predicting only a single quality metric lack interpretability in the results. Therefore, later methods aimed to disentangle different dimensions impacting audio quality. For example, the NISQA \cite{Mittag_2021_nisqa} method divides the prediction of speech quality into four dimensions: \textit{Noisiness, Coloration, Distortion, and Loudness}. The authors of the MOSRA framework \cite{hajal2022mosrajointmeanopinion} add room acoustic descriptors and SNR to their multi-task learning (MTL) framework enhancing interpretability. Finally, the DNSMOS \cite{reddy2022dnsmosp835nonintrusiveperceptual} model is one of the most commonly used methods that follow the ITU-T recommendation P.835 \cite{p835rec}, dividing the quality estimation of denoised speech into signal, background, and overall quality estimates.

These above MTL methods take a step toward general-purpose audio assessment by mimicking the human capability of rating different sources and dimensions of impairments in the audio signal. However, these methods do so in a static manner, and they cannot dynamically attend to different parts of the signal. Moreover, the above methods have all focused on speech-specific quality assessment.

We cast the audio assessment task as a ``text and audio to text'' prediction task to address these limitations. We call our method semi-intrusive as it does not compare against a clean reference signal but rather a description of the signal and characteristic to be assessed (e.g., speech, music, sneezing, MOS, SNR, etc.). Further, the choice of using textual output enables the model to predict a wide range of audio metrics, such as MOS and SNR, as well as tasks like distortion-type classification. Moreover, we hope that casting the task in this way paves the way for true explainability, where a model can directly describe and quantify audio impairments.

The proposed semi-intrusive method leverages instruction fine-tuning of the multi-modal PENGI model \cite{deshmukh2024pengiaudiolanguagemodel}, which has shown good performance on other audio-related tasks. We conduct experiments on both MOS prediction for speech and music, as well as SNR estimation for environmental sounds. Due to the lack of quality-labeled music datasets, we simulate data using an intrusive metric. For the SNR estimation of environmental sounds, we utilize the ESC-50 dataset \cite{piczak2015dataset}. The results demonstrate that our method achieves competitive performance for both speech and music quality MOS estimation. Furthermore, to our knowledge, we introduce the first end-to-end SNR estimator capable of predicting the SNR for arbitrary signal classes within a mixture of sounds. 

 
\begin{figure*}[t!]
    \centering
    \caption{Overview of the PENGI architecture adapted for semi-intrusive audio assessment. The encoders process the audio and text inputs, and together with the mapping networks $m1$ and $m2$, they generate the soft prompt for the GPT2-base language model. Components marked red are fine-tuned during training, while those marked blue have frozen parameters. The masked labels are fed as input to the model during training to enable training with teacher forcing.}
    \label{fig:pengi_architecture}
    \includegraphics[width=0.8\linewidth]{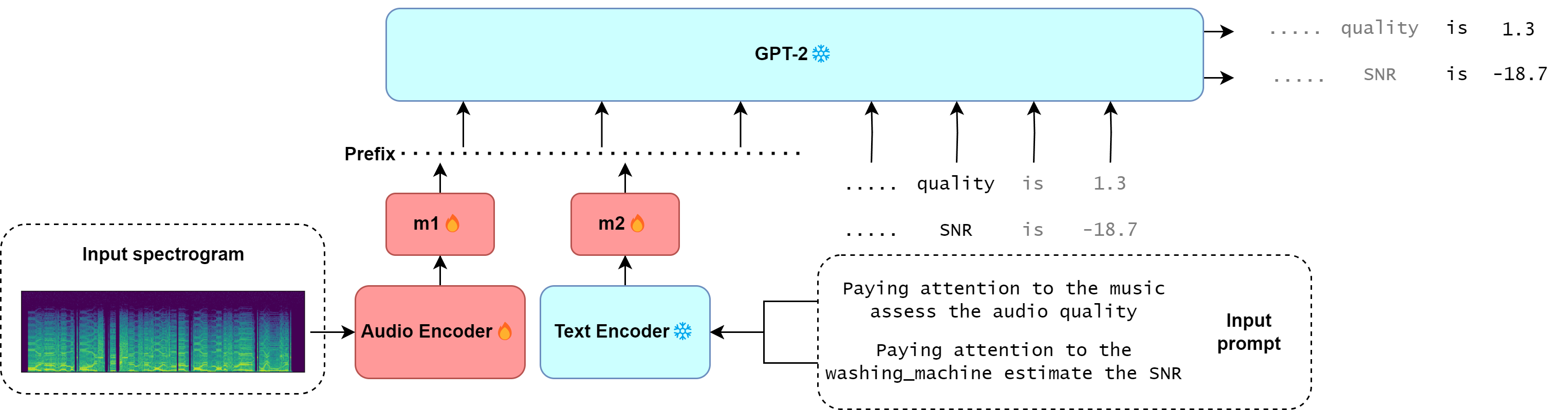}    
\end{figure*}

\section{Methods}

We present our method for semi-intrusive audio quality estimation, where the task is framed as a text prediction problem using both audio and text inputs, with the text providing a description of the signal class.

\subsection{Model architecture}
We use the PENGI architecture, a multi-modal model trained as a captioning system conditioned on the audio and the input prompt. The model has three main components: the audio encoder, text encoder, and causal language model. Additionally, two mapping networks translate the output embeddings of the audio and text encoders to the latent space of the causal language model. An overview of the model and its adaptation to semi-intrusive assessment is shown in \autoref{fig:pengi_architecture}. Note that the masked labels are only used as input to the language model during training to enable learning using teacher forcing. 

The PENGI model uses the HTS-AT \cite{chen2022htsathierarchicaltokensemanticaudio} model as audio encoder $a_\theta$, which maps a given audio signal $s \in \mathbb{R}^n$ into a fixed size vector $z_a = a_\theta (s) \in \mathbb{R}^{768}$. Following the implementation of the original authors, we resampled the input audio files to 44.1 kHz. The resulting files are then truncated/ randomly padded to a fixed length of 7 seconds. The model takes as input 64-band mel-spectrograms computed with a hop size of 10 ms and a window length of 32 ms. 

The text encoder computes a single text embedding from the input prompt. A transformer-based model is used in CLIP \cite{radford2021learningtransferablevisualmodels}. It consists of a stack of 12 transformer layers with a width 512 and 8 attention heads. The final representation of the input prompt is taken as the activation after the last transformer layer of the [EOS] token. Thus given an input prompt $q$, the text encoder $t_\xi$ maps it to a fixed size vector $z_t = t_\xi (q) \in \mathbb{R}^{512}$. During our fine-tuning, the text encoder remains frozen.

The audio and text encoders output a single embedding in a joint multi-modal space. To translate the embeddings into the latent space of the language model, two 8-layer transformers $m1$ and $m2$ are used to map a single embedding vector $z$ into a 40-length prefix $p = m(z) \in \mathbb{R}^{40 \times 768}$.

The base version of GPT-2 \cite{radford2019language} is used as the causal language model to generate text-based quality assessments. During fine-tuning, this model remains frozen. At inference time, the language model generates the quality assessment text autoregressively, conditioned on the prefix generated by the mapping networks. 

\subsection{Fine-tuning}

Our semi-intrusive model is obtained by supervised fine-tuning of the pre-trained PENGI model. To this end, we utilize the standard captioning loss. Given the audio prefix $p_a$ and the text prefix $p_t$ the $k-$length soft prompt for the causal language model is defined as $x = \text{Concat}\{p_a, p_t\}$. The model is then trained to predict the next token in the caption $c$ of length $l$ using cross entropy defined in \autoref{eq:caption_loss}.
\begin{equation}
    \mathcal{L} = - \sum^l_{i=1} \text{log }p(c_i | x_1, \dots, x_k, c_1, \dots, c_{i-1})
    \label{eq:caption_loss}
\end{equation}

\subsection{Prompting and labelling strategy}
In line with the PENGI model, we develop instruction templates for audio quality and SNR estimation. We hypothesize that conditioning the model based on the type of audio signal will allow the model to focus on relevant features within the audio. The instruction template, shown in \autoref{tab:templates}, includes the type of audio as part of its structure. Since we aim to model a regression task using a language model, we need to quantize the ground truth labels. In this work, we investigate two different quantization strategies for audio quality estimation.

\begin{table}[]
\scriptsize
\centering
\caption{Overview of the prompt and label template.}
\label{tab:templates}
\begin{tabular}{cc}
\toprule
Prompt                                                                                                             & Label                                                                                       \\ \midrule
\begin{tabular}[c]{@{}c@{}}Paying attention to the \{speech, music, ...\} \\ assess the audio quality\end{tabular} & \begin{tabular}[c]{@{}c@{}}The audio quality is \\ \{Quantized quality label\}
\end{tabular}\\
\bottomrule
\end{tabular}

\end{table}

\subsection{Data simulation}
\subsubsection{Quality estimation}
Due to the scarcity of non-speech datasets with human-annotated quality labels, we utilize data simulation with an intrusive metric to generate training pairs. In the simulation, we take a clean audio signal $s$ and apply an augmentation $a$ to create a distorted version. We then use an intrusive metric $f$ to produce training data \{$v,y$\} where $v=a(s)$ and $y=f(s,v)$.

\subsubsection{SNR estimation}
To show the generality of our model we build a semi-intrusive SNR estimator leveraging data simulation. For this, we take a dataset containing various audio classes and mix them at a specific SNR. Thus given two audio signals $s_i$ and $s_j$ containing different classes $\kappa_i$ and $\kappa_j$ we mix the signals at a specific SNR and ask the model to estimate the SNR of the signal containing the class $\kappa_i$.

\section{Experiments}
\subsection{Training details}
We fine-tune the PENGI model using the ADAMW optimizer \cite{loshchilov2019decoupledweightdecayregularization} with a fixed learning rate of 0.0001. We train the network for 15 epochs with a global batch size of 96 distributed across two RTX 4070 GPUs in half precision. For evaluation of the test sets, we select the checkpoint with the lowest validation loss.

\subsection{Label quantization strategy}
In this study, we evaluate two different strategies for quantizing the ground truth labels for our quality estimator. First, based on the recommendations from the p.800 listening test specifications, we quantize the MOS labels to the nearest integer and assign the corresponding textual description used in listening tests. Additionally, we compare this approach with a strategy where the labels are rounded to the first decimal place, using the string representation of the number as the label. \autoref{tab:quantization_strat} provides an overview of these label quantization strategies along with example labels.

For our SNR estimator, we use the quantization, where we round to the first decimal place. 

\subsection{Evaluation}
We generate quality captions through autoregressive generation during evaluation, conditioned on both the text and audio prefix. The model outputs discrete tokens that represent the assessed audio quality. We extract the numerical representation from the text according to the quantization strategy, e.g., "The audio quality is bad" $\rightarrow 1.0$.

We assess the model's performance using the Pearson correlation coefficient defined in \autoref{eq:pearsonr} for the audio quality estimation paradigm.
\begin{equation}
    r(x,y) = \dfrac{\Sigma_{i=1}^n (x_i - \overline{x})(y_i - \overline{y})}{\sqrt{\Sigma_{i=1}^n (x_i - \overline{x})^2} \sqrt{\Sigma_{i=1}^n (y_i - \overline{y})^2}}
    \label{eq:pearsonr}
\end{equation}

We evaluate the model using the root mean squared error defined in \autoref{eq:rmse} for the semi-intrusive SNR estimation.
\begin{equation}
    \text{RMSE}(x,y) = \sqrt{\dfrac{1}{N} \Sigma^N_{i=1} (x_i - y_i)^2}
    \label{eq:rmse}
\end{equation}

\begin{table}[]
\centering
\scriptsize
\caption{Overview of the quantization strategies and example labels.}
\label{tab:quantization_strat}
\begin{tabular}{ccc}
\toprule
Quality label & Text based               & Number based             \\ \midrule
1.3242        & Bad                      & 1.3                      \\
3.3345        & Fair                     & 3.3                      \\
4.798         & Excellent                & 4.8                      \\

\midrule
      Example label        & The audio quality is bad & The audio quality is 1.3 \\
              \bottomrule
\end{tabular}
\end{table}

\subsection{Datasets}
\subsubsection{Quality estimation}
Our semi-intrusive quality estimation network focuses on audio signals containing impairments typically found in teleconferencing and streaming applications. 

To fine-tune and evaluate our semi-intrusive audio quality estimator, we combine a large number of speech-focused human-labeled datasets. Our training dataset is composed of the Tencent corpus \cite{yi2022conferencingspeech2022challengenonintrusive}, the training portion of the NISQA dataset \cite{Mittag_2021_nisqa}, and the PSTN corpus \cite{mittag20b_interspeech_pstn}. For validation purposes, we use the validation sets of NISQA and experiments 1A, 1D, and 1O from the P-Supplement 23 dataset \cite{P.Sup23}. Finally, our testing datasets are composed of: the NISQA test datasets LIVETALK, FOR, and P501, the internal ReverbSpeechQuality dataset \cite{nessler21_interspeech}, the TCD-VoIP dataset \cite{harte2015tcd}, and experiments 3A/C/D/O from the P-Supplement 23 dataset.

We also assess our method using non-speech/mixed data. To this end, we leverage the ODAQ \cite{torcoli2023odaqopendatasetaudio} and Core SV \cite{coresv2014listening} datasets. These datasets contain speech and music signals degraded by various methods. For the Core SV dataset, the clean audio signals are processed by six lossy compression methods. The signals in the ODAQ dataset are distorted by
impairments related to coding artifacts such as filtering and pre-echos and impairments related to neural network processing. Both datasets contain scores obtained from a MUSHRA~\cite{BS_1534_3} listening test and, as such, have a scale of 0-100. We linearly map the MUSRA scores to the MOS absolute category rating scale between 1 and 5. Furthermore, we manually classify the signals in both datasets to obtain the signal class.

To obtain training and validation data for non-speech signals, we leverage data simulation with a focus on music signals. We utilize the small subset of the Free Music Archive (FMA) \cite{fma_dataset} for clean signals. For augmentations, we focus on distortions commonly associated with audio coding. The distortions applied include MP3 codec, Vorbis codec, µ-law encoding, Opus codec, and lowpass filtering. The parameters for these augmentations are chosen to ensure that the distortions are perceptually audible based on informal listening. An overview of the distortions is provided in \autoref{tab:semi_sim_distort}.

Per distortion, we sample 2000 clean signals for training and 500 for validation. They are resampled to 44.1 kHz, truncated to 7 seconds, and impaired using the sampled distortion. The parameters for the distortion are sampled uniformly, with the exception of the low-pass filter, which is sampled uniformly in the Mel scale. The clean and distorted audio signals are then used to generate the pseudo-label using the intrusive metric. 

To investigate which intrusive metrics are suitable for pseudo label generation, we evaluate them on the ODAQ dataset as it contains both traditional DSP-based distortions as well as distortions related to processing by neural-based methods. We consider the intrusive metrics: VISQOL \cite{chinen2020visqolv3opensource}, VISQOLAudio v3 \cite{visqolaudio}, HAAQI \cite{haaqi}, PESQ (wideband) \cite{pesq}, and POLQA (super wideband) \cite{beerends2013perceptual}. Since these methods operate at different sample rates, we resample the ODAQ signals to match the native sample rate for each metric.

\begin{table}[]
\centering
\scriptsize
\caption{Overview of the augmentations used for data simulation.}
\label{tab:semi_sim_distort}
\begin{tabular}{lcr}
\toprule
Augmentation       & Parameters              & Values                  \\ \midrule
MP3 Codec          & {[}8,16,24,32,48,56{]} & kbit/s                  \\
Vorbis Codec       & {[}-1, 1, 2{]}         & q-scale (quality scale) \\
$\mu$-law encoding & {[}24, 28, 32, 36{]}   & bits per sample         \\
Opus Codec         & {[}8, 12, 14, 16{]}    & kbit/s                  \\
Lowpass filtering  & 1000 - 6000            & Hz          \\
\bottomrule
\end{tabular}
\end{table}

\subsubsection{SNR Estimation}
For our semi-intrusive SNR estimator, we train and test on simulated data based on the ESC-50 dataset \cite{piczak2015dataset}. It contains 5-second-long signals in 50 different environmental classes, such as animal sounds and natural soundscapes. To generate a training pair $\{v, snr_c\}$, we sample two signals of different classes $s_c$ and $s_n$. These are then mixed at a specific SNR for the reference signal $s_c$. Given the different task, we construct an additional instruction template with the prompt being \textit{"Paying attention to the \{class\} estimate the SNR"}, and the label being \textit{"The SNR is \{$snr_c$\}"}.\\
We use the original splits from the ESC-50 dataset and thus generate training data from splits 1 to 3, validation data from split 4, and testing data from split 5. In total, we generate 72000 training, 4000 validation, and 4000 testing samples. 

\subsection{Baselines}
\subsubsection{Quality estimation}
To evaluate the effectiveness of our proposed framework for audio quality assessment, we compare it against two baselines: the supervised MOSRA model and the unsupervised PAM method. The MOSRA model \cite{hajal2022efficientspeechqualityassessment} uses XLS-R \cite{conneau2020unsupervisedcrosslingualrepresentationlearning} embeddings and trains a downstream network to estimate MOS scores. We compare our results with both the original MOSRA model and a version re-trained on our datasets to isolate the effects of the data and model architecture. The MOSRA model operates at a sample rate of 16 kHz, so all data is resampled accordingly.

The Prompting Audio Models (PAM) framework by \cite{deshmukh2024pampromptingaudiolanguagemodels} utilizes the CLAP model \cite{CLAP2023} for audio quality estimation. This method employs an antonym prompting strategy, where the audio embedding is compared to two text embeddings representing "good" and "bad" quality. The cosine distances between the audio embedding and the prompts are passed through a softmax to produce a probability distribution, with the quality score derived from the probability associated with the "good" prompt.\\

\subsubsection{SNR estimation}
To the best of our knowledge, a non-intrusive system for SNR estimation capable of focusing on specific sources in a mixture signal does not exist. As such, we set the baseline to be the same model without the signal class in the prompt, i.e., we prompt the model with \textit{"Paying attention to the audio estimate the SNR"} for all signals.

\section{Results and discussion}
\subsection{Quality estimation}

\autoref{tab:intr_met_res} shows the linear correlation coefficients of selected intrusive metrics on the ODAQ dataset. We find that HAAQI achieves the lowest correlation, followed by PESQ. The three top-performing metrics are all close regarding correlation, with POLQA being the best. However, due to the minimal difference between these metrics and the open-source nature of VISQOL Audio, we selected VISQOL Audio as the intrusive metric for pseudo-labeling in our data simulation pipeline.

\begin{table}[]
\scriptsize
\centering
\caption{Pearson correlation of the intrusive metrics on ODAQ.}
\label{tab:intr_met_res}
\begin{tabular}{l|ccccc}
\toprule
Dataset & VISQOL Audio v3 & VISQOL & HAAQI & PESQ & POLQA         \\ \midrule
ODAQ & 0.62         & 0.6    & 0.33  & 0.41 & \textbf{0.63}\\
\bottomrule
\end{tabular}
\end{table}

The left column of \autoref{tab:quality_res} shows the results of our model trained on both human-labeled and simulated datasets. Overall, our model achieves high correlation on the speech datasets with the exeption of the TCD-Chop and TCD-Compspkr datasets. For the mixed datasets we observe a general decrease in performance, particularly on the ODAQ dataset. When comparing label quantization strategies, we see improved performance with numeric label quantization across both speech and mixed datasets. The speech datasets show a small 0.03 difference in Pearson correlation, while the mixed datasets show a larger 0.1 absolute gap. On average, numeric quantization results in a 10.7\% relative improvement in performance.
\autoref{tab:quality_res} also shows the comparison with the baseline systems. The results show that our system outperforms the baselines on average with an average of 0.62. However, we find that the original MOSRA model performs the best on the speech datasets, with an average Pearson correlation of 0.89 compared to 0.79 for our method. The original MOSRA model, however, fails on the mixed datasets, achieving random performance. Re-training MOSRA on our data improves its performance on ODAQ to 0.39 but also slightly decreases its performance on the speech datasets. Nonetheless, the re-trained MOSRA model is unable to predict the quality for the Core SV dataset. 

\begin{table}[]
\scriptsize
\centering
\caption{Results of the semi-intrusive MOS estimator on the test datasets. On the left the results of our method are shown for the two different label quantization strategies. On the right the baseline methods MOSRA and PAM. Presented scores are Pearson correlation.}
\label{tab:quality_res}
\begin{tabular}{lccccc}
\toprule
                      & \multicolumn{2}{c}{Ours}                    & \multicolumn{2}{c}{MOSRA \cite{hajal2022efficientspeechqualityassessment}}                   & PAM \cite{deshmukh2024pampromptingaudiolanguagemodels}                 \\ \cline{2-6} 
                      & text                 & numeric              & Original             & Retrained            & -                    \\ \hline
ReverbSpeechQuality (s)  & 0.86                 & 0.90                 & 0.92                 & 0.89                 & 0.69                 \\
NISQA FOR (s)     & 0.81                 & 0.86                 & 0.91                 & 0.89                 & 0.32                 \\
NISQA LIVETALK (s) & 0.78                 & 0.84                 & 0.91                 & 0.9                  & 0.42                 \\
NISQA P501 (s)     & 0.87                 & 0.91                 & 0.93                 & 0.91                 & 0.33                 \\
P Suppl 23-3A (s)    & 0.72                 & 0.70                 & 0.82                 & 0.8                  & 0.44                 \\
P Suppl 23-3C (s)   & 0.76                 & 0.79                 & 0.93                 & 0.91                 & 0.58                 \\
P Suppl 23-3D (s)   & 0.70                 & 0.70                 & 0.89                 & 0.9                  & 0.27                 \\
P Suppl 23-3O (s)   & 0.76                 & 0.80                 & 0.91                 & 0.86                 & 0.51                 \\
TCD-Chop (s)             & 0.59                 & 0.61                 & 0.86                 & 0.79                 & 0.14                 \\
TCD-Clip (s)             & 0.88                 & 0.87                 & 0.83                 & 0.8                  & 0.52                 \\
TCD-Compspkr (s)         & 0.60                 & 0.62                 & 0.87                 & 0.85                 & 0.14                 \\
TCD-Echo (s)             & 0.72                 & 0.73                 & 0.88                 & 0.85                 & 0.43                 \\
TCD-Noise (s)             & 0.84                 & 0.90                 & 0.89                 & 0.84                 & 0.54                 \\
Core SV (m)               & 0.48                 & 0.59                 & 0.04                 & 0.08                 & 0.47                 \\
ODAQ (m)                  & 0.23                 & 0.30                 & 0.03                 & 0.39                 & 0.4                  \\ \midrule
Speech average (s)           & 0.76                 & 0.79                 & \textbf{0.89}        & 0.86                 & 0.41                 \\
Mixed average (m)         & 0.35                 & \textbf{0.45}        & 0.04                 & 0.24                 & 0.44                 \\
Average               & 0.56                 & \textbf{0.62}        & 0.46                 & 0.55                 & 0.42                \\
\bottomrule
\end{tabular}

\end{table}

The results also point out that the PAM method performs poorly on the speech datasets with an average correlation of 0.41, and, similarly to our model performs comparatively worse on TCD-Chop and TCD-Compspkr. However, PAM achieves the highest performance on ODAQ at 0.4, and good correlation on Core SV (0.39). Despite the good results on the mixed datasets, it achieves a lower average correlation than the other methods.

Additionally, we conducted an ablation study on both the inclusion of the simulated data and the fine-tuning of the PENGI audio encoder which showed that both additions were beneficial for the overall performance of the system for MOS prediction. Due to space constraints, we chose to omit the complete analysis which can be found online \cite{coldenhoff2024objective}.

\subsection{SNR estimation}
The results for our semi-intrusive SNR estimator are in \autoref{tab:snr_res}. The table shows that the model with the constant prompt struggles to consistently predict the SNR for the selected signal class. In particular, when compared to the random baseline, the model without the signal class included in the prompt performs at a random level. However, when the signal class is included in the prompt, our semi-intrusive SNR estimator significantly improves, achieving an RMSE of 7.5, compared to 16.56 for the random baseline. These results demonstrate that including the signal class in the prompt allows the model to focus on specific sources in the mixture, significantly improving SNR estimation.

\begin{table}[]
\scriptsize
\centering
\caption{RMSE (dB) results for our semi-intrusive SNR estimator on the simulated ESC-50 dataset. }
\label{tab:snr_res}
\begin{tabular}{lccc}
\toprule
Sound categories                                                              & Semi-intrusive & Fixed prompt & Random \\ \midrule
Animals                                                                       & 6.73           & 15.4        & 16.48  \\
\begin{tabular}[c]{@{}c@{}}Natural soundscapes \&\\ water sounds\end{tabular} & 7.48           & 15.71        & 16.53  \\
Human, non-speech                                                             & 8.33           & 15.66        & 16.53  \\
Interior/domestic sounds                                                      & 7.19           & 16.18        & 16.61  \\
Exterior/urban noises                                                         & 7.69          & 15.65        & 16.55  \\ 
\midrule
All                                                                           & 7.5           & 15.72       & 16.56 \\
\bottomrule
\end{tabular}

\end{table}

\section{Conclusion}

In this work, we presented a novel semi-intrusive approach to audio assessment, designed to mimic the human ability to focus on different components of an audio signal. By framing the task as a text prediction problem conditioned on both audio and text inputs, our method eliminates the need for a clean reference signal and leverages the flexibility of textual descriptions for a wide range of audio evaluation tasks.

Our experiments demonstrate that the proposed approach achieves competitive performance in MOS estimation for both speech and music, outperforming baseline methods on average. 
Additionally, we introduced our semi-intrusive SNR estimator capable of predicting the SNR for arbitrary signal classes within a mixture, showing that the model can selectively attend to specific sources in complex audio mixtures. We aim for our proposed semi-intrusive method to take a step into the direction of general-purpose audio assessment. Future work could expand this approach by incorporating more non-speech datasets and extending the method to broader audio evaluation tasks using textual descriptions and detailed quantifications of audio impairments.

\newpage

\balance

\bibliographystyle{IEEEbib}
\bibliography{refs}

\begin{thebibliography}{10}

\bibitem{p800rec}
ITU-T~Recommendation P.800,
\newblock ``{Methods for objective and subjective assessment of quality},''
  1996.

\bibitem{BS_1284}
ITU-R~Recommendation BS.1284,
\newblock ``{General methods for the subjective assessment of sound quality},''
  2003.

\bibitem{Dimolitsas_1989}
S.~Dimolitsas,
\newblock ``{Objective speech distortion measures and their relevance to speech
  quality assessments},''
\newblock in {\em IEE Proceedings I (Communications, Speech and Vision)}, 1989,
  pp. 317--324.

\bibitem{chinen2020visqolv3opensource}
Michael Chinen, Felicia S.~C. Lim, Jan Skoglund, Nikita Gureev, Feargus
  O'Gorman, and Andrew Hines,
\newblock ``{ViSQOL v3\: An Open Source Production Ready Objective Speech and
  Audio Metric},''
\newblock in {\em Twelfth International Conference on Quality of Multimedia
  Experience (QoMEX)}, 2020.

\bibitem{pesq}
A.W. Rix, J.G. Beerends, M.P. Hollier, and A.P. Hekstra,
\newblock ``{Perceptual evaluation of speech quality (PESQ)-a new method for
  speech quality assessment of telephone networks and codecs},''
\newblock in {\em ICASSP 2001 - 2001 IEEE International Conference on
  Acoustics, Speech and Signal Processing (ICASSP)}, 2001, pp. 749--752.

\bibitem{beerends2013perceptual}
John~G Beerends, Christian Schmidmer, Jens Berger, Matthias Obermann, Raphael
  Ullmann, Joachim Pomy, and Michael Keyhl,
\newblock ``{Perceptual objective listening quality assessment (POLQA), the
  third generation {ITU-T} standard for end-to-end speech quality measurement
  part i—temporal alignment},''
\newblock in {\em Journal of the audio engineering society}, 2013, pp.
  366--384.

\bibitem{kumar2023torchaudiosquimreferencelessspeechquality}
Anurag Kumar, Ke~Tan, Zhaoheng Ni, Pranay Manocha, Xiaohui Zhang, Ethan
  Henderson, and Buye Xu,
\newblock ``{Torchaudio-Squim: Reference-Less Speech Quality and
  Intelligibility Measures in Torchaudio},''
\newblock in {\em ICASSP 2023 - 2023 IEEE International Conference on
  Acoustics, Speech and Signal Processing (ICASSP)}, 2023.

\bibitem{patton2016automoslearningnonintrusiveassessor}
Brian Patton, Yannis Agiomyrgiannakis, Michael Terry, Kevin Wilson, Rif~A.
  Saurous, and D.~Sculley,
\newblock ``{AutoMOS: Learning a non-intrusive assessor of
  naturalness-of-speech},''
\newblock in {\em NIPS 2016 End-to-end Learning for Speech and Audio Processing
  Workshop}, 2016.

\bibitem{fu2018qualitynetendtoendnonintrusivespeech}
Szu wei Fu, Yu~Tsao, Hsin-Te Hwang, and Hsin-Min Wang,
\newblock ``{Quality-Net: An End-to-End Non-intrusive Speech Quality Assessment
  Model Based on BLSTM},''
\newblock in {\em Proc. Interspeech 2018}, 2018, pp. 1873--1877.

\bibitem{Mittag_2021_nisqa}
Gabriel Mittag, Babak Naderi, Assmaa Chehadi, and Sebastian Möller,
\newblock ``{NISQA: A Deep CNN-Self-Attention Model for Multidimensional Speech
  Quality Prediction with Crowdsourced Datasets},''
\newblock in {\em Proc. Interspeech 2021}, 2021, pp. 2127--2131.

\bibitem{hajal2022mosrajointmeanopinion}
Karl~El Hajal, Milos Cernak, and Pablo Mainar,
\newblock ``{MOSRA: Joint Mean Opinion Score and Room Acoustics Speech Quality
  Assessment},''
\newblock in {\em Proc. Interspeech 2022}, 2022, pp. 3313--3317.

\bibitem{reddy2022dnsmosp835nonintrusiveperceptual}
Chandan K~A Reddy, Vishak Gopal, and Ross Cutler,
\newblock ``{DNSMOS P.835: A Non-Intrusive Perceptual Objective Speech Quality
  Metric to Evaluate Noise Suppressors},''
\newblock in {\em ICASSP 2022 - 2022 IEEE International Conference on
  Acoustics, Speech and Signal Processing (ICASSP)}, 2022, pp. 886--890.

\bibitem{p835rec}
ITU-T~Recommendation P.835,
\newblock ``{Subjective test methodology for evaluating speech communication
  systems that include noise suppression algorithm},'' 2003.

\bibitem{deshmukh2024pengiaudiolanguagemodel}
Soham Deshmukh, Benjamin Elizalde, Rita Singh, and Huaming Wang,
\newblock ``{Pengi: An Audio Language Model for Audio Tasks},''
\newblock in {\em Advances in Neural Information Processing Systems}, 2023, pp.
  18090--18108.

\bibitem{piczak2015dataset}
Karol~J. Piczak,
\newblock ``{ESC: Dataset for Environmental Sound Classification},''
\newblock in {\em Proc. of the 23rd Annual ACM Conference on Multimedia}, 2015,
  pp. 1015--1018.

\bibitem{chen2022htsathierarchicaltokensemanticaudio}
Ke~Chen, Xingjian Du, Bilei Zhu, Zejun Ma, Taylor Berg-Kirkpatrick, and Shlomo
  Dubnov,
\newblock ``{HTS-AT: A Hierarchical Token-Semantic Audio Transformer for Sound
  Classification and Detection},''
\newblock in {\em ICASSP 2022 - 2022 IEEE International Conference on
  Acoustics, Speech and Signal Processing (ICASSP)}, 2022, pp. 646--650.

\bibitem{radford2021learningtransferablevisualmodels}
Alec Radford, Jong~Wook Kim, Chris Hallacy, Aditya Ramesh, Gabriel Goh,
  Sandhini Agarwal, Girish Sastry, Amanda Askell, Pamela Mishkin, Jack Clark,
  Gretchen Krueger, and Ilya Sutskever,
\newblock ``{Learning Transferable Visual Models From Natural Language
  Supervision},''
\newblock in {\em International Conference on Machine Learning}, 2021, pp.
  8748--8763.

\bibitem{radford2019language}
Alec Radford, Jeff Wu, Rewon Child, David Luan, Dario Amodei, and Ilya
  Sutskever,
\newblock ``{Language Models are Unsupervised Multitask Learners},''
\newblock 2019.

\bibitem{loshchilov2019decoupledweightdecayregularization}
Ilya Loshchilov and Frank Hutter,
\newblock ``{Decoupled Weight Decay Regularization},''
\newblock in {\em International Conference on Learning Representations}, 2017.

\bibitem{yi2022conferencingspeech2022challengenonintrusive}
Gaoxiong Yi, Wei Xiao, Yiming Xiao, Babak Naderi, Sebastian Möller, Wafaa
  Wardah, Gabriel Mittag, Ross Cutler, Zhuohuang Zhang, Donald~S. Williamson,
  Fei Chen, Fuzheng Yang, and Shidong Shang,
\newblock ``{ConferencingSpeech 2022 Challenge: Non-intrusive Objective Speech
  Quality Assessment (NISQA) Challenge for Online Conferencing Applications},''
\newblock in {\em Proc. Interspeech 2022}, 2022, pp. 3308--3312.

\bibitem{mittag20b_interspeech_pstn}
Gabriel Mittag, Ross Cutler, Yasaman Hosseinkashi, Michael Revow, Sriram
  Srinivasan, Naglakshmi Chande, and Robert Aichner,
\newblock ``{DNN No-Reference PSTN Speech Quality Prediction},''
\newblock in {\em Proc. Interspeech 2020}, 2020, pp. 2867--2871.

\bibitem{P.Sup23}
ITU-T~Recommendation P.Sup23,
\newblock ``{ITU-T coded-speech database},'' 1998.

\bibitem{nessler21_interspeech}
Natalia Nessler, Milos Cernak, Paolo Prandoni, and Pablo Mainar,
\newblock ``{Non-Intrusive Speech Quality Assessment with Transfer Learning and
  Subject-Specific Scaling},''
\newblock in {\em Proc. Interspeech 2021}, 2021, pp. 2406--2410.

\bibitem{harte2015tcd}
Naomi Harte, Eoin Gillen, and Andrew Hines,
\newblock ``{TCD-VoIP, a research database of degraded speech for assessing
  quality in VoIP applications},''
\newblock in {\em 2015 Seventh International Workshop on Quality of Multimedia
  Experience (QoMEX)}. IEEE, 2015.

\bibitem{torcoli2023odaqopendatasetaudio}
Matteo Torcoli, Chih-Wei Wu, Sascha Dick, Phillip~A. Williams, Mhd
  Modar~Halimeh, William Wolcott, and Emanuël A.~P. Habets,
\newblock ``{ODAQ: Open Dataset of Audio Quality},''
\newblock in {\em ICASSP 2024 - 2024 IEEE International Conference on
  Acoustics, Speech and Signal Processing (ICASSP)}, 2024, pp. 836--840.

\bibitem{coresv2014listening}
{CoreSV Team},
\newblock ``{CoreSV listening test},'' 2014.

\bibitem{BS_1534_3}
ITU-R~Recommendation BS.1534-3,
\newblock ``{Method for the subjective assessment of intermediate quality level
  of audio systems},'' 2015.

\bibitem{fma_dataset}
Micha\"el Defferrard, Kirell Benzi, Pierre Vandergheynst, and Xavier Bresson,
\newblock ``{FMA: A Dataset for Music Analysis},''
\newblock in {\em 18th International Society for Music Information Retrieval
  Conference (ISMIR)}, 2017.

\bibitem{visqolaudio}
Colm Sloan, Naomi Harte, Damien Kelly, Anil~C. Kokaram, and Andrew Hines,
\newblock ``{Objective Assessment of Perceptual Audio Quality Using
  ViSQOLAudio},''
\newblock {\em IEEE Transactions on Broadcasting}, pp. 693--705, 2017.

\bibitem{haaqi}
James~M. Kates and Kathryn~H. Arehart,
\newblock ``{The Hearing-Aid Audio Quality Index (HAAQI)},''
\newblock {\em IEEE/ACM Transactions on Audio, Speech, and Language
  Processing}, pp. 354--365, 2016.

\bibitem{hajal2022efficientspeechqualityassessment}
Karl El~Hajal, Zihan Wu, Neil Scheidwasser-Clow, Gasser Elbanna, and Milos
  Cernak,
\newblock ``{Efficient Speech Quality Assessment Using Self-Supervised
  Framewise Embeddings},''
\newblock in {\em ICASSP 2023 - 2023 IEEE International Conference on
  Acoustics, Speech and Signal Processing (ICASSP)}, 2023.

\bibitem{conneau2020unsupervisedcrosslingualrepresentationlearning}
Alexis Conneau, Alexei Baevski, Ronan Collobert, Abdelrahman Mohamed, and
  Michael Auli,
\newblock ``{Unsupervised Cross-Lingual Representation Learning for Speech
  Recognition},''
\newblock in {\em Proc. Interspeech 2021}, 2021, pp. 2426--2430.

\bibitem{deshmukh2024pampromptingaudiolanguagemodels}
Soham Deshmukh, Dareen Alharthi, Benjamin Elizalde, Hannes Gamper, Mahmoud {Al
  Ismail}, Rita Singh, Bhiksha Raj, and Huaming Wang,
\newblock ``{PAM: Prompting Audio-Language Models for Audio Quality
  Assessment},''
\newblock in {\em Proc. Interspeech 2024}, 2024, pp. 3320--3324.

\bibitem{CLAP2023}
Benjamin Elizalde, Soham Deshmukh, and Huaming Wang,
\newblock ``{Natural Language Supervision For General-Purpose Audio
  Representations},''
\newblock in {\em ICASSP 2024 - 2024 IEEE International Conference on
  Acoustics, Speech and Signal Processing (ICASSP)}, 2024, pp. 336--340.

\bibitem{coldenhoff2024objective}
Jozef Coldenhoff,
\newblock ``Objective perception metrics for audio quality,''
\newblock in {\em EPFL Infoscience}, 2024, pp. 35--36.

\end{thebibliography}

\end{document}